\documentclass[12pt,showpacs,showkeys,amsmath,amssymb]{revtex4}
\usepackage{amsmath,amsfonts,amsthm,amscd,amssymb,latexsym}
\usepackage{bm}
\usepackage{dcolumn}
\usepackage{graphicx}
\usepackage{epstopdf}
\usepackage{color}
\usepackage{epsf}
\usepackage{epsfig}
\usepackage{graphicx, epic, eepic, color}
\usepackage{soul}
\usepackage[colorlinks=true,urlcolor=blue,linkcolor=blue,citecolor=blue]{hyperref}

\newcommand{\beq}{\begin{equation}}
\newcommand{\eeq}{\end{equation}}
\newcommand{\bea}{\begin{eqnarray}}
\newcommand{\eea}{\end{eqnarray}}

\begin{document}

\title{Gravitomagnetic  Helicity}

\author{Donato Bini$^{1,2}$}
\email{donato.bini@gmail.com}
\author{Bahram Mashhoon$^{3,4}$}
\email{mashhoonb@missouri.edu}
\author{Yuri N. Obukhov$^5$}
\email{obukhov@ibrae.ac.ru}
\affiliation{}

  \affiliation{
$^1$Istituto per le Applicazioni del Calcolo ``M. Picone", \\
 CNR, I-00185 Rome, Italy\\
$^2$INFN, Sezione di Roma Tre, I-00146 Rome, Italy\\
$^3$School of Astronomy,
Institute for Research in Fundamental Sciences (IPM),
P. O. Box 19395-5531, Tehran, Iran\\
$^4$Department of Physics and Astronomy,
University of Missouri, Columbia,
Missouri 65211, USA\\
$^5$Theoretical Physics Laboratory, Nuclear Safety Institute, \\
Russian Academy of Sciences, B.Tulskaya 52, 115191 Moscow, Russia}

\date{\today}
\begin{abstract}
Mass currents in astrophysics generate gravitomagnetic fields of enormous complexity. Gravitomagnetic helicity, in direct analogy with magnetic helicity, is a measure of entwining of the gravitomagnetic field lines. We discuss gravitomagnetic helicity within the gravitoelectromagnetic (GEM) framework of linearized general relativity. Furthermore, we employ the spacetime curvature approach to GEM in order to determine the gravitomagnetic helicity for static observers in Kerr spacetime.  
\end{abstract}


\maketitle

\section{Introduction}\label{Intro}

In fluid mechanics, helicity is defined as $\bm{v}\cdot \bm{\omega}$, where $\bm{v}$ is the fluid velocity and $\bm{\omega} = \bm{\nabla} \times \bm{v}$ is the vorticity. Helicity is a measure of the intertwisting of vortices in the flow of the fluid. Total helicity is the integral of helicity over a given volume of space. In a similar way, magnetic helicity is defined to be $\bm{A}\cdot \bm{B}$, where $\bm{A}$ is the electromagnetic vector potential and $\bm{B} = \bm{\nabla} \times \bm{A}$ is the magnetic field~\cite{HeOb}. Magnetic helicity is a measure of the twisting and linking of the magnetic field lines~\cite{Elsasser, Woltjer, Vishniac:2000br, Blackman:2014kxa, Kedia:2016nwk, Sche}.

This classical notion of magnetic helicity is quite different from the standard quantum notion of the helicity of a particle that has to do with its spin. However, by combining the magnetic helicity with the dual object known as the electric helicity, one can introduce the electromagnetic (or optical) helicity that is directly related to the helicity of light understood as the projection of photon's angular momentum on the direction of motion \cite{Trueba,Cameron,Bliokh,Fernandez-Corbaton:2020shg}. 

The purpose of this paper is to extend the classical notion of magnetic helicity to general relativity and the gravitational  analog of the magnetic field that has to do with the current of \emph{matter}. Previous work in this direction is contained in~\cite{Afanasiev,Alves:2018wku}. Throughout this paper, Greek indices run from $0$ to $3$, while Latin indices run from $1$ to $3$; moreover,  the signature of the spacetime metric is $+2$ in our convention. In Section \ref{GEM}, we carefully review the approximate nature of the analogy with electromagnetism within the context of linearized general relativity. These are old topics that go back to Einstein's work~\cite{Einstein}. The physical interpretation of the resulting formalism is the subject of Section \ref{Matter}. Gravitomagnetic helicity is discussed in Sections \ref{Helicity} and \ref{Curvature}. Finally, Section \ref{Discussion} contains a brief discussion of our results.

\section{Gravitomagnetic Field}\label{GEM}

Electric currents generate magnetic fields; similarly, mass currents are expected to generate gravitomagnetic fields. These are of fundamental interest as they represent non-Newtonian aspects of the gravitational field. The gravitomagnetic field of the Earth has been measured in the Gravity Probe B (GP-B) test of general relativity~\cite{Everitt:2011hp, Everitt:2015qri}. In linearized general relativity (GR), the framework of gravitoelectromagnetism (GEM) has been quite useful in elucidating the gravitational effects of rotating masses; in this connection, see~\cite{BCT, Mash93, Mashhoon:2003ax, Mashhoon:2000he, Harris, jfps, mlrtart1, mlrtart2, Bini:2008cy, Costa:2021atq} and the references cited therein.  Let us therefore consider a rotating astronomical body that is confined to a compact region of space surrounding the origin of the spatial Cartesian coordinates~\cite{Poisson}.  We work in the weak-field and slow-motion approximation of GR. In this case, the spacetime metric in Cartesian spacetime coordinates $x^\mu = (ct, x, y, z)$  can be expressed as
\begin{equation}\label{M1}
 g_{\mu \nu} = \eta_{\mu \nu} + h_{\mu \nu}\,, 
\end{equation}
where  $h_{\mu \nu}$ is the linear symmetric gravitational perturbation of the Minkowski spacetime metric tensor $\eta_{\mu \nu}=\,$diag$(-1, 1, 1, 1)$, in our convention. It proves useful to introduce the trace-reversed potentials
\begin{equation}\label{M2}
\bar{h}_{\mu \nu}=h_{\mu \nu}-\frac{1}{2}\eta_{\mu \nu}h\,, \qquad
h:=\eta_{\mu \nu}h^{\mu \nu}\,, \qquad \bar{h}=-h\,.
\end{equation} 
The ten gravitational potentials are gauge dependent.  Under an infinitesimal coordinate transformation, 
\begin{equation}\label{M3}
 x^\mu \mapsto x'^\mu=x^\mu-\varepsilon^\mu(x)\,, 
\end{equation}
the linear change in the gravitational potentials is given by 
\begin{equation}\label{M4}
\bar{h}\,'_{\mu \nu}=\bar{h}_{\mu \nu}+\varepsilon_{\mu,\nu}+\varepsilon_{\nu,\mu}-\eta_{\mu \nu}
\,\varepsilon^\alpha{}_{,\alpha}\,, \quad \bar{h}\,'=\bar{h}-2\,\varepsilon^\alpha{}_{,\alpha}\,.
\end{equation}
As expected, the linearized gravitational field as well as the corresponding field equations remains invariant under gauge transformations; that is, the Riemann tensor remains invariant, but the connection changes. 

Einstein's gravitational field equation is given by
\begin{equation}\label{M5}
 R_{\mu \nu} - \frac{1}{2} g_{\mu \nu}R = \frac{8 \pi G}{c^4} T_{\mu \nu}\,,
\end{equation}
where $T_{\mu \nu}$ is the conserved,  $T^{\mu \nu}{}_{;\mu} = 0$, symmetric energy-momentum tensor of matter and we have set the cosmological constant equal to zero ($\Lambda = 0$). 
The linearized gravitational field equation can be expressed as 
\begin{equation}\label{M6}
-\frac{1}{2}\,\Box\, \bar{h}_{\mu \nu}+\bar{h}^\rho{}_{(\mu,\nu)\rho}
-\frac{1}{2}\eta_{\mu \nu}\bar{h}^{\rho \sigma}{}_{,\rho \sigma} = \frac{8 \pi G}{c^4} T_{\mu \nu}\,,
\end{equation}
where $\Box :=\eta^{\alpha \beta}\partial_\alpha \partial_\beta$ and  $T^{\mu \nu}{}_{,\mu} = 0$ . 

It proves convenient to impose the transverse gauge condition
\begin{equation}\label{M7}
\bar{h}^{\mu\nu}{}_{, \nu}=0\,.
\end{equation}
The gauge is not quite fixed, however, since a gauge transformation such that $\Box\, \varepsilon_\mu = 0$ is still possible. 

With the imposition of  gauge condition~\eqref{M7}, the gravitational field equations of linearized GR take the form
\begin{equation}\label{M8}
\Box\, \bar{h}_{\mu \nu} = -\,{\frac{16 \pi G}{c^4}}T_{\mu \nu}\,.  
\end{equation}
The linear superposition of the general solution of the source-free wave equation and a particular source-dependent solution constitutes the general solution  of Eq.~\eqref{M8}. The particular solution can be a linear combination of the retarded and advanced solutions.  We are interested in the field that is purely generated by the gravitational source under consideration here; therefore, we choose the particular retarded solution 
\begin{equation}\label{M9} 
\bar{h}_{\mu\nu} = \frac{4G}{c^4}\int\frac{T_{\mu\nu}(ct-|\bm{x}-\bm{x}'|,\bm{x}')}{|\bm{x}-\bm{x}'|}\,d^3x'\,,
\end{equation}
where $(-4\pi\,|\bm{x}-\bm{x}'|)^{-1}$ is the Green function for the three-dimensional Laplace operator.   We assume the source consists of slowly moving matter ($|\bm{v}| \ll c$) of density $\rho$, pressure $p$ and matter current $\bm{j} = \rho\,\bm{v}$. The matter energy-momentum tensor can be approximated by  a perfect fluid and written as
\begin{equation}\label{M10}
T_{\mu \nu} \approx \rho\,c^2\, u_\mu u_\nu + p (\eta_{\mu \nu} +  u_\mu u_\nu)\,, \qquad u^\mu = \frac{dx^\mu}{d\tau} \approx \left(1, \frac{\bm{v}}{c}\right)\,.  
\end{equation} 
Thus, in terms of the dominant expressions in powers of $c$, we have  $T_{00} \approx \rho c^2$, $T_{0i} \approx -\,c\,j_i$ and $T_{ij} \approx \rho v_iv_j + p\delta_{ij}$. The corresponding gravitational potentials are $\bar{h}_{00} = -\,4\Phi /c^2$, $\bar{h}_{0i} = -\,2A_i/c^2$ and $\bar{h}_{ij} = O(c^{-4})$.  We neglect $O(c^{-4})$ terms in the gravitational potentials.  Under these conditions, the spacetime
metric has the GEM form
\begin{equation}\label{M11} 
ds^2 = -\,c^2\left(1+2\frac{\Phi}{c^2}\right)dt^2-\frac{4}{c}(\bm{ A}\cdot d\bm{x})dt+\left(1-2\frac{\Phi}{c^2}\right) \delta_{ij}dx^idx^j\,,
\end{equation}
where 
\begin{equation}\label{M12} 
\Phi (t, \bm{x}) = -\,G\int\frac{\rho(ct-|\bm{x}-\bm{x}'|,\bm{x}')}{|\bm{x}-\bm{x}'|}\,d^3x'\,, \quad \bm{A} (t, \bm{x}) = \frac{2G}{c}\int\frac{\bm{j}(ct-|\bm{x}-\bm{x}'|,\bm{x}')}{|\bm{x}-\bm{x}'|}\,d^3x'\,.
\end{equation}
In the Newtonian regime, the gravitoelectric potential $\Phi$ reduces to the Newtonian gravitational potential, while the gravitomagnetic potential $\bm{A}$ vanishes, since $\bm{A}=O(c^{-1})$. 

The transverse gauge condition~\eqref{M7} can be written as
\begin{equation}\label{M13}
\bar{h}^{00}{}_{, 0} + \bar{h}^{0i}{}_{, i} = 0\,, \qquad \bar{h}^{i0}{}_{, 0} + \bar{h}^{ij}{}_{, j} = 0\,.
\end{equation}
In terms of the GEM potentials, we have
\begin{equation}\label{M14}
-\,{\frac{1}{c}}\frac{\partial \Phi}{\partial t}+\bm{\nabla} \cdot \left(\frac{1}{2}\bm{A}\right)=0\,,
\qquad  \frac{2}{c^3}\frac{\partial A^i}{\partial t} + \bar{h}^{ij}{}_{, j} = 0\,.
\end{equation}
Here, $ \bar{h}^{ij}{}_{, j}$ vanishes because $\bar{h}_{ij} = O(c^{-4})$ and we have neglected $O(c^{-4})$ terms in the gravitational potentials. Furthermore,  we learn from Eq.~\eqref{M12} that the gravitomagnetic vector potential $\bm{A}$ is non-Newtonian and of order $1/c$ or smaller. Hence, 
\begin{equation}\label{M15} 
\bar{h}^{i0}{}_{,0} = \frac{\partial}{\partial(ct)}\left(\frac{2A^i}{c^2}\right) = \frac{2}{c^3} \frac{\partial A^i}{\partial t}\,,
\end{equation}
where $A^i = O(1/c)$. No matter what value the quantity $\partial A^i/\partial t$ takes, $\bar{h}^{i0}{}_{,0}$ is still of order  $1/c^4$ and therefore negligible in our approximation scheme. One cannot logically conclude anything about   $\partial A^i/\partial t$ from the transverse gauge condition. Some authors, e.g.~\cite{ciufowhee},  deal with stationary situations from the outset and thus assume $\partial \bm{A}/\partial t = 0$. In some other contexts, the theory imposes this condition as in the extension of the GEM formalism to nonlocal gravity~\cite{Mashhoon:2019jkq}.  
 
We \emph{define} the GEM fields via
\begin{equation}\label{M16} 
\bm{E} = \bm{\nabla} \Phi -\frac{1}{c}\frac{\partial}{\partial t}\left(\frac{1}{2}\bm{A}\right),\quad \bm{B}=\bm{\nabla} \times \bm{A}\,,
\end{equation}
in direct analogy with electromagnetism. Here, we employ the convention described in detail in~\cite{Mash93, Mashhoon:2003ax, Mashhoon:2000he}. It follows from these definitions that the GEM fields have dimensions of acceleration and
\begin{equation}\label{M17} 
\bm{\nabla} \times \bm{E} = -\,{\frac{1}{c}}\frac{\partial }{\partial t}\left(\frac{1}{2}
\bm{B}\right)\,,\quad \bm{\nabla} \cdot \left(\frac{1}{2}\bm{B}\right)=0\,.
\end{equation}
Furthermore, Eq.~\eqref{M8} implies
\begin{equation}\label{M18} 
\bm{\nabla}\cdot\bm{E} = 4\pi G\rho\,, \qquad \bm{\nabla} \times \left(\frac{1}{2}\bm{B}\right)
= {\frac{1}{c}}\frac{\partial}{\partial t} \bm{E}+\frac{4\pi G}{c}\bm{j}\,.
\end{equation}
As expected, these equations contain the continuity equation 
\begin{equation}\label{M19} 
 \bm{\nabla}\cdot \bm{j}+\frac{\partial \rho}{\partial t} =0\,.
\end{equation}
The scalar and vector operations throughout refer to Euclidean space conventions. The extra factor of $1/2$ in the GEM field equations originates from the fact that the underlying linearized general  relativity theory involves a spin-2 field.  

If the source distribution is \emph{stationary} and confined around the origin of spatial coordinates, then far from the source
\begin{equation}\label{M20} 
\Phi \sim -\,{\frac{GM}{|\bm{x}|}},\quad \bm{A}\sim \frac{G}{c}\frac{\bm{J}\times \bm{x}}{|\bm{x}|^3}\,,
\end{equation}
where $M$ and $\bm{J}$ are the mass and angular momentum of the source, respectively.  In a steady state situation, $\rho$ and $\bm{j}$ are independent of time and we have  $\bm{\nabla} \cdot \bm{j} = 0$, $\bm{\nabla} \cdot \bm{A} = 0$ and
\begin{equation}\label{M21} 
\Phi(\bm{x}) = -\,G\int\frac{\rho(\bm{y})}{|\bm{x}-\bm{y}|}\,d^3y\,, \qquad
\bm{A} (\bm{x}) = \frac{2G}{c}\int\frac{\bm{j}(\bm{y})}{|\bm{x}-\bm{y}|}\,d^3y\,.
\end{equation}
Assuming $|\bm{x}| > |\bm{y}|$, which is valid for the exterior of the source, and expanding $|\bm{x} - \bm{y}|$ to first order in $|\bm{y}|/|\bm{x}|$, we get the Newtonian expression for $\Phi$ to lowest order. On the other hand, for the vector potential we have
\begin{equation}\label{M22}
\int\frac{\bm{j}(\bm{y})}{|\bm{x} - \bm{y}|}\,d^3y \approx \frac{1}{|\bm{x}|}\, \int\bm{j}(\bm{y})\,d^3y
+\frac{1}{|\bm{x}|^3}\, \int (\bm{x} \cdot \bm{y})\,\bm{j}(\bm{y})\,d^3y\,.
\end{equation}
Consider a finite closed spatial domain $\mathcal{D}$ that completely surrounds the  compact gravitational  source such that $\bm{j}$ vanishes on the surface of $\mathcal{D}$ and beyond. The conservation of matter current $\bm{\nabla} \cdot \bm{j} = 0$ implies
\begin{equation}\label{M23}
\int_{\mathcal{D}} \mathcal{I}(\bm{y})\, (\bm{x} \cdot \bm{y})\,
\bm{\nabla}_{\bm{y}} \cdot \bm{j}(\bm{y})\,d^3y = 0\,,
\end{equation}
where $\mathcal{I}(\bm{y})$ is a smooth function.
Employing Gauss's theorem and setting the boundary integral equal to zero, we find
\begin{equation}\label{M24}
\int_{\mathcal{D}} \bm{\nabla}_{\bm{y}}[\mathcal{I}(\bm{y})\,
(\bm{x} \cdot \bm{y})]\cdot \bm{j}(\bm{y})\,d^3y = 0\,.
\end{equation}
For $\mathcal{I}(\bm{y}) = 1$ and $\mathcal{I}(\bm{y}) = y^i$, we get the following relations
\begin{equation}\label{M25}
\int_{\mathcal{D}} \bm{j}(\bm{y})\,d^3y = 0\,,\qquad \int_{\mathcal{D}} (\bm{x}\cdot\bm{y})
\,j^i(\bm{y})\,d^3y = - \int_{\mathcal{D}} y^i\, [\bm{x} \cdot \bm{j}(\bm{y})]\,d^3y\,,
\end{equation}
respectively. The total proper \emph{angular momentum} of the gravitational source is given by
\begin{equation}\label{M26}
\bm{J} = \int_{\mathcal{D}} \bm{y} \times \bm{j}(\bm{y})\,d^3y\,.
\end{equation}
Therefore, 
\begin{equation}\label{M27}
\bm{J} \times \bm{x} = \int_{\mathcal{D}}(\bm{x} \cdot \bm{y})\, \bm{j}(\bm{y})\,d^3y
- \int_{\mathcal{D}} \bm{y} \,[\bm{x} \cdot \bm{j}(\bm{y})]\,d^3y\,.
\end{equation}
This relation together with Eq.~\eqref{M25} implies
\begin{equation}\label{M28}
\frac{1}{2}\, \bm{J} \times \bm{x} = \int_{\mathcal{D}} (\bm{x} \cdot \bm{y})\,\bm{j}(\bm{y})\,d^3y\,.
\end{equation}
Putting the above results together, we conclude that far from the source the gravitomagnetic vector potential to lowest order  is given by
\begin{equation}\label{M29}
\bm{A}(\bm{x}) = \frac{G}{c}\,\frac{\bm{J} \times \bm{x}}{|\bm{x}|^3}\,
\end{equation}
and the corresponding gravitomagnetic field can be expressed as  
\begin{equation}\label{M30}
\bm{B}(\bm{x}) = \frac{G}{c}\,\frac{3\,(\bm{J}\cdot \bm{x})\,\bm{x} - \bm{J}\,|\bm{x}|^2}{|\bm{x}|^5}\,. 
\end{equation}

It is interesting to discuss the observational status of the non-Newtonian gravitomagnetic field in the solar system. It turns out that the gravitomagnetic field of the Earth has been directly measured via the GP-B experiment and the GR prediction has been verified to about 19\%~\cite{Everitt:2011hp, Everitt:2015qri}. The internal orbital angular momentum of the  Earth-Moon system  acts as an extended gyroscope in the GEM field of the Sun~\cite{Desitter, Fokker, Shapiro}; moreover, the gravitomagnetic field of the Sun affects the Earth-Moon distance.  In connection with the lunar laser ranging experiment, we note that the main relativistic effects in the motion of the Moon are due to the gravitational field of the Sun and have been calculated in Refs.~\cite{MaTh1, MaTh2}.
 
For the sake of completeness, we need a GEM analog of the Lorentz force law. We can write the Lagrangian, $L=-\,mc\, ds/dt$, for the motion of a free test particle of mass $m$ to linear order in $\Phi$ and $\bm{A}$ as
\begin{equation}\label{M31} 
L = -\,{\frac{mc^2}{\gamma}} - m\gamma\left(1+\frac{v^2}{c^2}\right) \Phi
- {\frac{2m}{c}}\gamma\,\bm{v}\cdot \bm{A}\,,
\end{equation}
where $\gamma$ is the Lorentz factor. The geodesic equation takes on a familiar form if $\partial \bm{A}/\partial t=0$; then, the equation of motion for the kinetic momentum $\bm{p}=\gamma m\bm{v}$ becomes $d\bm{p}/dt= \bm{F}$, and to lowest order in $\bm{v}/c$, $\Phi$ and $\bm{A}$, the force $\bm{F}$ can be written as
\begin{equation}\label{M32}
\bm{F} = -\,m\bm{E} - 2m\frac{\bm{v}}{c}\times \bm{B}\,.
\end{equation}
In this case, $\bm{P} = \bm{p} + (-2m/c)\bm{A}$ is the canonical momentum of the particle. The influence of the gravitomagnetic field on orbital motion can be illustrated via the Lense-Thirring effect; in this connection,  see Refs.~\cite{MaHeT,ILRC, Renz} and the references cited therein.

Let us now discuss the remaining gauge freedom of the GEM potentials. We recall from the discussion following Eq.~\eqref{M7} that the transverse gauge condition is maintained if $\varepsilon^\mu$ is such that   $\Box \varepsilon^\mu=0$. Working \emph{exclusively with the trace-reversed potentials}, we can choose $\varepsilon^\mu$ in Eq.~\eqref{M4} in such a manner that the GEM potentials transform as in electrodynamics and at the same time those elements of the connection defining GEM fields remain invariant. To this end, let $\varepsilon_0=O(c^{-3})$ and $\varepsilon_i=O(c^{-4})$; then, Eq.~\eqref{M4} implies that
\begin{equation}\label{M33}
\Phi '= \Phi +\frac{1}{c}\frac{\partial }{\partial t}\Psi,\quad
{\frac{1}{2}}\bm{A}' = {\frac{1}{2}}\bm{A} + \bm{\nabla}\Psi\,, 
\end{equation}
where $\Psi =c^2\varepsilon ^0/4$ and $\Box \Psi=0$. Under the gauge transformation~\eqref{M33}, the GEM fields are invariant in close analogy with electrodynamics. In particular, with $\Psi = -\,ct \Phi_0 -\tfrac{1}{2}\bm{x} \cdot \bm{A}_0$, where $\Phi_0$ and $\bm{A}_0$ are infinitesimal constants, we can shift the GEM potentials by constant amounts such that $(\Phi, \bm{A}) \mapsto (\Phi - \Phi_0, \bm{A}-\bm{A}_0)$.

The basic linear perturbation theory described above must be supplemented with the gravitational Larmor theorem for its proper interpretation~\cite{Mash93}. Historically, Larmor's theorem established a local correspondence between magnetism and rotation. The gravitational Larmor theorem is essentially Einstein's principle of equivalence within the GEM framework. Furthermore, there is a close correspondence between GEM and electromagnetism. The GEM approach has been useful for experimental gravitation~\cite{BCT, ciufowhee, Tar1}  and the method  has been applied to a variety of situations~\cite{Ruggiero:2020nef, Ruggiero:2020oxo, Mashhoon:2021qtc}. Henceforth, we will denote the  gravitoelectric and gravitomagnetic fields by $\bm{E}_g$ and $\bm{B}_g$ to distinguish them from the corresponding electromagnetic fields.  

The rotation of a mass affects the spacetime structure around it. The gravitomagnetic temporal structure has been elucidated via the gravitomagnetic clock effects~\cite{CoMa, MGT, MGL, BJM1, BJM2, Hackmann:2014aga}, the gravitomagnetic time delay~\cite{CKMR}, etc. On the other hand, $c^2/B_g$ has dimensions of length and the gravitomagnetic spatial structure  can be demonstrated by studying propagation of waves in a gravitomagnetic field, which  is the subject of the next section.

\section{Matter Waves in a Gravitomagnetic Field}\label{Matter}

The non-Newtonian gravitomagnetic field has been measured~\cite{Everitt:2011hp, Everitt:2015qri} and is therefore important for fundamental physics. We illustrate the application of GEM approach in this section via an example. In electrodynamics, the magnetic vector potential and the constant magnetic field inside a rotating spherical shell of uniform charge density with its center at the origin of Cartesian coordinates are given by
\begin{equation}\label{L1}
\bm{A} = \frac{1}{2} \bm{B} \times \bm{x}\,, \qquad  \bm{B} = \frac{2}{3} \frac{Q_0}{cR_0}\,\bm{\Omega}_0\,,
\end{equation}
where $Q_0$ is the total charge, $R_0$ is the radius  and $\bm{\Omega}_0$ is the constant rotation frequency of the shell. The electric field vanishes throughout the interior of the shell. In the GEM weak-field and slow-motion approximation of GR, inside the corresponding \emph{slowly rotating} uniform shell of matter we have~\cite{MaHeT}
\begin{equation}\label{L2}
\bm{A}_g = \frac{1}{2} \bm{B}_g \times \bm{x}\,, \qquad  \bm{B}_g = \frac{4GM_0}{3cR_0} \,\bm{\Omega}_0\,,\qquad  \bm{E}_g = 0\,,
\end{equation}   
where $M_0$ is the total mass of the electrically neutral shell. 

In our coordinate system $x^\mu = (ct, x, y, z)$, we assume that the massive spherical shell surrounding the origin of spatial coordinates slowly rotates about the $z$ axis; then, the spacetime metric inside the shell can be written in GEM form as 
\begin{equation}\label{L3}
ds^2 = g_{\mu \nu}\, dx^\mu dx^\nu =  \eta_{\mu \nu}\, dx^\mu dx^\nu + 2 \frac{B_g}{c} dt\,(ydx-xdy)\,,
\end{equation}
where $B_g = 4GM_0\Omega_0/(3cR_0)$ is constant and we have set the constant Newtonian potential inside the shell equal to zero via a gauge transformation as explained in Section \ref{GEM}. Let us now introduce cylindrical coordinates $(\varrho, \varphi, z)$ in this stationary spacetime, namely,
\begin{equation}\label{L4}
x = \varrho \cos \varphi\,, \qquad y = \varrho \sin\varphi\,;
\end{equation}
then, to linear order the transformed metric  takes the form
\begin{equation}\label{L5}
ds^2 = -\,c^2 dt^2 -2\frac{B_g}{c} \varrho^2 dt d\varphi +d\varrho^2 + \varrho^2 d\varphi^2 + dz^2\,.
\end{equation}
To linear order in $B_g/c$, this metric is equivalent to the Minkowski spacetime metric in a system of coordinates that rotates about the $z$ axis with angular velocity $B_g/c$. This is in conformity with the gravitational Larmor theorem. To describe the propagation of waves in this spacetime, one could in principle employ the transformation properties of the corresponding fields. However, we adopt a more direct approach in what follows.  

\subsection{Scalar field}\label{scalar}

Consider next a scalar field ${\mathit \Phi}$ of inertial mass $m$ in our interior spacetime,
\begin{equation}\label{L6}
g^{\mu \nu} {\mathit \Phi}_{; \mu \nu} - \frac{m^2c^2}{\hbar^2} {\mathit \Phi} = 0\,,
\end{equation}
where ${\frac \hbar {mc}}$ is the Compton wavelength of the particle. The scalar wave equation can be written as 
\begin{equation}\label{L7}
\frac{1}{\sqrt{-g}}\,\frac{\partial}{\partial x^\mu} \left(\sqrt{-g} \,g^{\mu \nu}\frac{\partial {\mathit \Phi}}{\partial x^\nu} \right)  - \frac{m^2c^2}{\hbar^2} {\mathit \Phi} = 0\,.
\end{equation}
In our case, $\sqrt{-g} = \varrho$ to linear order and the wave equation reduces to 
\begin{equation}\label{L8}
 -\frac{1}{c^2}\frac{\partial^2 {\mathit \Phi}}{\partial t^2}  -2\frac{B_g}{c^3}\,\frac{\partial^2 {\mathit \Phi}}{\partial t \partial \varphi} +\frac{1}{\varrho}\frac{\partial}{\partial \varrho}\left(\varrho \frac{\partial {\mathit \Phi}}{\partial \varrho}\right) 
+\frac{1}{\varrho^2}\frac{\partial^2 {\mathit \Phi}}{\partial \varphi^2} +\frac{\partial^2 {\mathit \Phi}}{\partial z^2} - \frac{m^2c^2}{\hbar^2} {\mathit \Phi} = 0\,.
\end{equation}
The spacetime metric is invariant under translations in $t$, $\varphi$ and $z$; hence, we can assume a solution of the form
\begin{equation}\label{L9}
 {\mathit \Phi} = e^{-i \omega t + i \mu \varphi +i k_z z} f(\varrho)\,,
\end{equation}
where $\omega$, $\mu$ and $k_z$ are constants. The stationary configuration is invariant under $\varphi \mapsto \varphi + 2 \pi$; therefore, we conclude that $\mu = n =  0, \pm 1, \pm 2, \pm 3,\cdots$. Then, Eq.~\eqref{L8} implies
\begin{equation}\label{L10}
\omega^2 = {\frac{2 nB_g \omega}{c}} +  \omega_0^2\,,
\end{equation}
where $\omega_0$ is the free wave frequency in absence of the gravitomagnetic field,
\begin{equation}\label{om0}
\omega_0^2 = {\frac{m^2c^4}{\hbar^2}} +  c^2k_z^2 + c^2 \eta^2\,,
\end{equation}
and we assume $\eta > 0$. Moreover, the radial part of the wave equation reduces to the differential equation
\begin{equation}\label{L11}
\varrho^2 \frac{d^2f}{d\varrho^2} + \varrho \frac{df}{d\varrho} + (\eta^2 \varrho^2 - n^2)f(\varrho) = 0\,.
\end{equation}
Using $\eta \varrho$ as the new dimensionless independent variable, we recover the Bessel equation. We are interested in solutions that are nonsingular at $\varrho = 0$; therefore, $f = J_n (\eta \varrho)$, where $J_n$ is the Bessel function of the first kind~\cite{A+S}.  We recall that for a Bessel function of order $n$,  $J_{-n} (x) = (-1)^n J_n(x)$. 

The scalar wave propagation is influenced by the gravitomagnetic field in the specific manner that appears in Eq.~\eqref{L10}. One way to look at Eq.~\eqref{L10} is to write its positive-frequency solution to linear order in $B_g$  as
\begin{equation}\label{L12}
\omega = \omega_0 + \,n\frac{B_g}{c}\,. 
\end{equation}
That is, the free wave frequency $\omega_0$ becomes $\omega_0 \,+\, n B_g/c$ in the presence of the gravitomagnetic field, which indicates that $B_g/c$ has the interpretation of gravitomagnetic frequency here.   

On the other hand, on the basis of Eq.~\eqref{L10}, one can assign a certain wave number to $(B_g \omega/c^3)^{1/2}$ and a corresponding wavelength to $[c^3/(B_g \omega)]^{1/2}$. If we replace $\omega$ here by the Compton frequency of the particle $mc^2/\hbar$, we obtain the \emph{gravitomagnetic length} $\ell$ that is defined in analogy with the magnetic case via
\begin{equation}\label{L13}
 \ell  := \left(\frac{\hbar c}{mB_g}\right)^{1/2}\,.
\end{equation}
Here, $\ell$ is essentially the smallest radius of a circular orbit for a particle of mass $m$ in the gravitomagnetic field $B_g$. To see this, note that it follows from Eq.~\eqref{M32} that the radius of a circular orbit of a particle around the gravitomagnetic field is $\varrho_0 = c v/(2B_g)$. The momentum of the particle is expected to satisfy an order-of-magnitude relation of the form  $mv \gtrsim \hbar/\varrho_0$ via the uncertainty principle. Hence, we conclude that $\varrho_0 \gtrsim \ell$.

\subsection{Spinor field}\label{spinor}

The dynamics of a spinor field ${\mathit \Psi}$ with inertial mass $m$ is determined by the covariant Dirac equation
\begin{equation}\label{Dirac0}
(i\hbar\gamma^{\hat \alpha} e^{\mu}{}_{\hat \alpha} D_\mu - mc){\mathit \Psi} = 0\,. 
\end{equation}
Here, $\gamma^{\hat \alpha}$ are the flat Dirac matrices, and for the spacetime geometry (\ref{L5}) the explicit form of the orthonormal frame $e^{\mu}{}_{\hat \alpha}$ and the covariant spinor derivative $D_\mu$ can be found in \cite{HN}. As a result, we recast the Dirac equation into the Schr\"odinger form
\begin{equation}
i\hbar\frac{\partial {\mathit \Psi}} {\partial t} = {\cal H}{\mathit \Psi}\,,\label{Sch} 
\end{equation}
where the Hermitian Hamiltonian reads explicitly:
\begin{equation}\label{Ham}
{\cal H} = \beta mc^2 + c\,\bm{\alpha}\cdot\widehat{\bm{p}} - {\frac{B_g}{c}}\,\widehat{L}{}_z\,.
\end{equation}
Here, $\widehat{\bm{p}} = -\,i\hbar\bm{\nabla}$ is the momentum operator and $\widehat{L}{}_z = \hbar\left(-\,i\partial_\varphi + {\frac 12}\,\Sigma_z\right)$ is the $z$ component of the total angular momentum along the direction of the external gravitomagnetic field. The Dirac matrices $\beta, \bm{\alpha}, \bm{\Sigma}$ are listed in Appendix~\ref{appdirac}.

By making use of an ansatz similar to (\ref{L9}), 
\begin{equation}\label{psi}
{\mathit \Psi}{}_{\pm} = e^{ -\,i\omega t + i\mu \varphi + i k_z z}\,S_{\pm}^{(s)}(\varrho, \varphi)\,,
\end{equation}
we find  $\mu = n =  0, \pm 1, \pm 2, \pm 3,\cdots$, and the frequencies now additionally depend on the spin variable $s = \pm 1$ (``up''/``down'' orientation):
\begin{equation}\label{ompm}
\omega = \omega_0 + \left(n + {\frac s2}\right){\frac{B_g}{c}}\,.
\end{equation}
Here, $\omega_0$ has the same form (\ref{om0}) as for the scalar field. The exact structure of the spinor amplitude $S_{\pm}^{(s)}$ is presented in Appendix~\ref{appdirac}. 

The frequency formula (\ref{ompm}), describing the effect of the gravitomagnetic field on matter waves, generalizes relation (\ref{L12}) to the case of nontrivial spin. The resulting spin-gravitomagnetic coupling is in general agreement with the treatment of Dirac equation in the exterior gravitational fields of rotating masses. Indeed, the gravitational field of the slowly rotating shell can be matched smoothly to a vacuum exterior that far from the source has the Lense-Thirring form~\eqref{M11}.  In such gravitational fields, the solution of the Dirac equation implies the presence of the same spin-gravitomagnetic field coupling as in Eq.~\eqref{ompm}, see~\cite{DeOT, BaMa, Papini:2007gx, Obukhov:2013zca, Obukhov:2017avp} and the references cited therein. 

The spin-gravitomagnetic coupling illustrates a general effect that could in principle be measured in a laboratory on the Earth~\cite{Mashhoon:2003ax, BaMa}; that is,  the difference between the energies of a Dirac particle with spin up and spin down in the gravitomagnetic field of the Earth is $\hbar B_g/c \sim 10^{-29}$ eV, which is still about seven orders of magnitude away from what can be presently measured in an earthbound laboratory~\cite{nEDM:2020crw}.

\section{Gravitomagnetic Helicity}\label{Helicity}

In astrophysics,  currents of charged particles on large scales generate magnetic fields of enormous complexities. As the relevant charged particles (electrons, positrons, protons, etc.) have masses, attendant gravitomagnetic fields are also generated with structures that are different from magnetic fields due to the universality of the gravitational interaction and hence its independence from electric charge. The notion of magnetic helicity has been introduced as a measure of the complexity of the magnetic field. In magnetohydrodynamics, the concept of magnetic helicity has been quite useful.  In this section, we explore the concept of gravitomagnetic helicity within the GEM formalism.

In analogy with the definition of magnetic helicity~\cite{Elsasser, Woltjer}  
\begin{equation}\label{0}
\mathbb{H} = \int \bm{A}\cdot\bm{B}\,d^3x=\int \bm{A}\cdot(\bm{\nabla}\times\bm{A})\,d^3x\,,
\end{equation}
we find it useful to define \emph{gravitomagnetic helicity} $\mathbb{H}_g$,
\begin{equation}\label{1}
\mathbb{H}_g = \int \bm{A}_g\cdot\bm{B}_g\,d^3x\,.
\end{equation}

Let us first note that $\mathbb{H}_g$ is a gauge invariant quantity. In fact, replacing $\bm{A}_g$ by $\bm{A}_g + \bm{\nabla} \Psi$, we find
\begin{equation}\label{2}
\int \bm{\nabla} \Psi \cdot \bm{B}_g\,d^3x  = \int [\bm{\nabla}\cdot(\Psi\bm{B}_g)
- \Psi \bm{\nabla} \cdot \bm{B}_g]\,d^3x = 0\,,
\end{equation}
by Gauss's theorem and $\bm{\nabla} \cdot \bm{B}_g = 0$. Moreover, the GEM field equations imply
\begin{equation}\label{3}
\partial_t \mathbb{H}_g =  -\,4c\,\int \bm{E}_g\cdot \bm{B}_g\, d^3x\,,
\end{equation}
which is completely gauge invariant as well. To show this, we start from
\begin{equation}\label{4}
\partial_t \mathbb{H}_g = \int [(\partial_t \bm{A}_g)\cdot \bm{B}_g
+ \bm{A}_g \cdot (\partial_t \bm{B}_g)]\,d^3x\,.
\end{equation}
From the GEM equations
\begin{equation}\label{5}
\partial_t {\bm A}_g = 2c\,[-\bm{E}_g + \bm{\nabla} \Phi]\,,\qquad
\partial_t \bm{B}_g = -\,2c\, \bm{\nabla} \times \bm{E}_g\,
\end{equation}
and 
\begin{equation}\label{6}
\bm{\nabla} \Phi \cdot \bm{B}_g = \bm{\nabla} \cdot (\Phi \bm{B}_g)\,, \qquad \bm{A}_g \cdot
(\bm{\nabla} \times \bm{E}_g) = -\,\bm{\nabla} \cdot(\bm{A}_g \times \bm{E}_g) +\bm{B}_g\cdot \bm{E}_g\,,
\end{equation}
one finds
\begin{equation}\label{7}
\partial_t \mathbb{H}_g = 2c\,\int [-\,2\bm{E}_g\cdot \bm{B}_g + \bm{\nabla}
\cdot (\Phi \bm{B}_g +\bm{A}_g \times \bm{E}_g)]\,d^3x\,,
\end{equation}
from which Eq.~\eqref{3} follows via Gauss's theorem.

What about \emph{gravitoelectric helicity}? In electrodynamics, free electromagnetic fields have basic physical significance; indeed, their existence implies the concept of electric helicity as well. An appropriate superposition of electric helicity and magnetic helicity leads to the notion of optical helicity, optical chirality and related phenomena~\cite{Bliokh,Afanasiev}. The linear perturbation approach to general relativity presented in Sections II and III has limited physical applicability within the full nonlinear framework of GR. In particular,  free GEM fields are devoid of any particular physical significance. Therefore, gravitoelectric helicity does not appear to be a meaningful concept in GEM.

\subsection{Topological aspects}\label{Top}

In electromagnetism (and in particular in magnetohydrodynamics), the magnetic helicity (\ref{0}) contains information about the topological structure of magnetic field lines and measures how they are linked, twisted and knotted~\cite{Elsasser, Woltjer, Vishniac:2000br, Blackman:2014kxa, Kedia:2016nwk, Sche}. 

On the other hand, the magnetic helicity can be naturally identified with the zeroth component $\bm{A}\cdot\bm{B} = -\,K^0$ of the Abelian Chern-Simons current $K^\mu = \epsilon^{\mu\nu\alpha\beta}A_\nu F_{\alpha\beta}$, the divergence of which is the density of the Chern topological charge,
\begin{equation}\label{dK}
\partial_\mu K^\mu = F_{\alpha\beta}\tilde{F}{}^{\alpha\beta} = 2\,\bm{E}\cdot\bm{B}\,,\qquad
\tilde{F}{}^{\alpha\beta} = {\frac 12}\,\epsilon^{\alpha\beta\mu\nu}F_{\mu\nu}\,.
\end{equation}
By integrating this over the volume, we verify that the total magnetic helicity $\partial_t\mathbb{H} = -\,2c\int\bm{E}\cdot\bm{B}\,d^3x$ is conserved for the electromagnetic field configurations with $\bm{E}\cdot\bm{B} = 0$.

Chern and Chern-Simons characteristic classes \cite{EGH} are important for the description of the topology of classical gauge fields and they both emerge in the context of the Atiyah-Singer theorem for the index of Dirac operator, explaining the structure of the quantum anomalies of the axial vector current. This phenomenon, when the classical symmetry and the corresponding Noether conservation law are broken by quantum effects, is universal in the sense that it arises for both fermion and boson axial currents. In the presence of an external gravitational field, the axial anomaly acquires additional contributions known as the Pontryagin \cite{EGH} and the Nieh-Yan \cite{Nieh:1982,Nieh:2018,Hehl:1991} topological terms. The latter is nontrivial on a spacetime with torsion \cite{Obukhov:1982,Obukhov:1983,Yajima:1985,Chandia,Obukhov:1997}. 

In order to put the discussion of the gravitomagnetic helicity into a proper topological framework, it is worthwhile to recall that Einstein's GR admits a reformulation in terms of the coframe variable \cite{reader,AP}, when the tetrad field $e^{\hat \alpha}{}_\mu$ is treated as the gravitational field potential and the torsion 
\begin{equation}\label{tor}
{\mathcal T}{}_{\mu\nu}{}^{\hat \alpha} = \partial_\mu e^{\hat \alpha}{}_\nu - \partial_\nu e^{\hat \alpha}{}_\mu
\end{equation}
arises as the corresponding gravitational field strength, in complete analogy with Maxwell's tensor $F_{\mu\nu} = \partial_\mu A_\nu - \partial_\nu A_\mu$. This approach has proved to be extremely useful for the development of the nonlocal extension of the gravitational theory \cite{nonlocal}.

By describing the GEM spacetime (\ref{M11}) in terms of the coframe 
\begin{equation}\label{cofGEM}
e^{\hat 0}{}_0 = 1 + {\frac 1{c^2}}\,\Phi\,,\qquad e^{\hat 0}{}_i = {\frac 2{c^2}}\,A_i\,,\qquad
e^{\hat i}{}_j = \delta^i_j\left(1 - {\frac 1{c^2}}\,\Phi\right)\,,  
\end{equation}
we then find the Nieh-Yan topological density  
\begin{equation}\label{NYT}
{\mathcal T}{}_{\mu\nu}{}^{\hat \alpha}\,\tilde{\mathcal T}{}^{\mu\nu}{}_{\hat \alpha} = -\,{\frac {16}{c^4}}
\,\bm{E}_g\cdot\bm{B}_g\,,\qquad \tilde{\mathcal T}{}^{\rho\sigma}{}_{\hat \alpha} =
{\frac 12}\,\epsilon^{\rho\sigma\mu\nu}{\mathcal T}_{\mu\nu}{}^{\hat \beta}\,g_{{\hat \alpha}{\hat \beta}}\,.
\end{equation}
As a result, the related Chern-Simons Nieh-Yan current reads:
\begin{equation}\label{KNY}
K_{\rm NY}^\rho = \epsilon^{\rho\mu\nu\lambda}e^{\hat \alpha}{}_\mu\,{\mathcal T}_{\nu\lambda}{}^{\hat \beta}\,g_{{\hat \alpha}{\hat \beta}}\,,
\qquad \partial_\rho K_{\rm NY}^\rho = {\mathcal T}{}_{\mu\nu}{}^{\hat \alpha}\,\tilde{\mathcal T}{}^{\mu\nu}{}_{\hat \alpha}\,,
\end{equation}
and its zeroth component,
\begin{equation}
K_{\rm NY}^0 = {\frac {4}{c^4}}\,\bm{A}_g\cdot\bm{B}_g\,,
\end{equation}
obviously determines the gravitomagnetic helicity
\begin{equation}\label{gemH}
\mathbb{H}_g = {\frac {c^4}{4}}\int K_{\rm NY}^0\,d^3x\,,
\end{equation}
which coincides with (\ref{1}). By integrating (\ref{KNY}) over the volume, we recover (\ref{3}). 

Currently, the study of the possible macroscopic manifestations of quantum anomalies in collective dynamics of systems of chiral particles is a new rapidly developing field of research \cite{871,kh1,kh2,app}. The chiral magnetic effect and the chiral vortical effect are the physically most interesting anomalous transport phenomena, and the torsional gravitational effects induced by the Nieh-Yan contribution to the axial anomaly attract special attention \cite{Volovik,Nissinen,Khaidukov,Imaki,Castillo:2015,CKS,Ferreiros}. 

The concept of gravitomagnetic helicity is expected to play a role in the complex physical phenomena associated with the merger of two Kerr black holes. The gravitational radiation emitted in this process could possibly carry away orbital angular momentum (OAM)~\cite{Baral}. The theoretical existence of linearized gravitational waves with OAM has recently been demonstrated~\cite{Iwo} by extending the construction of the knotted gravitational configurations, namely, the gravitational hopfions~\cite{Thompson1,Thompson2,Thompson3}. 

\subsection{An Example}\label{Ex}

In some simple cases described below, $\mathbb{H}_g$ vanishes. The GEM fields are linear; therefore, we can in principle consider a superposition of a number of fields. For instance, let
\begin{equation}\label{S1}
 \bm{A}_g = \bar{\bm{A}}_g +  \hat{\bm{A}}_g\,, \qquad   \bm{B}_g =  \bar{\bm{B}}_g +  \hat{\bm{B}}_g\, 
\end{equation}
and assume that $\bar{\mathbb{H}}_g = \hat{\mathbb{H}}_g = 0$. Then,  
\begin{equation}\label{S2}
\mathbb{H}_g   =  \int (\bar{\bm{A}}_g  \cdot \hat{\bm{B}}_g + \hat{\bm{A}}_g  \cdot \bar{\bm{B}}_g) \,d^3x\,.
\end{equation}

Let us briefly digress here and mention that the superposition $\bm{A}_g + \bm{A}'_g$ involving a \emph{constant} vector potential  $\bm{A}'_g$ with the corresponding $\bm{B}'_g = 0$ does not change  $\mathbb{H}_g$, since $\bm{A}'_g\cdot \bm{B}_g = -\,\bm{\nabla} \cdot (\bm{A}_g' \times \bm{A}_g)$, which upon integration over all space vanishes by Gauss's theorem.

Next, imagine a gravitomagnetic dipole ($\bm{J}/c$) at the origin of spatial coordinates with its field given by Eq.~\eqref{3}. We need to generalize this case; therefore,  consider the following pair of $(\bar{\bm{A}}_g, \bar{\bm{B}}_g)$ given by
\begin{equation}\label{S3}
 \bar{\bm{A}}_g = \frac{G}{c} \frac{\bm{J} \times  \bm{r} }{(r+\zeta)^n}\,, 
\end{equation}
\begin{equation}\label{S4}
 \bar{\bm{B}}_g =  \frac{G}{c}\, \frac{ n (\bm{J} \cdot  \bm{r})  \bm{r} +   \bm{J}\, r[2(r+\zeta) -nr] }{r(r+\zeta)^{n+1}}\,.
\end{equation}
This is a generalization of the standard dipolar field~\eqref{M30} for $n = 3, 4, 5, \cdots$, where $\zeta>0$ moderates the $r=0$ singularity. Here, $\bm{r} = (x, y, z)$ and $r = |\bm{r}|$; moreover, $\bar{\mathbb{H}}_g = 0$. 

We now need a second pair $(\hat{\bm{A}}_g, \hat{\bm{B}}_g)$ such that the magnetic field lines in this case are interlinked with those of ``dipolar" field lines. To this end, we consider a steady uniform  current of mass in the form of a line along the $z$ direction from $z = - \infty$ to $z = \infty$ given in Cartesian $(x, y, z)$ coordinates by
\begin{equation}\label{S5}
\hat{\bm{A}}_g =  \frac{4GI_m}{c}(0, 0, -\tfrac{1}{2}\ln (x^2+y^2))\,,
\end{equation}
\begin{equation}\label{S6}
 \hat{\bm{B}}_g = \frac{4GI_m}{c}\, \frac{1}{(x^2+y^2)} (-y, x, 0)\,,
\end{equation}
where $I_m$ is a constant and $\hat{\mathbb{H}}_g = 0$. Here, $I_m$ is the mass current, i.e. the amount of mass per unit time that moves up the $z$ axis,  in close analogy with the corresponding situation in electrodynamics, namely, 
\begin{equation}\label{S7}
 I_m = \int \bm{j} \cdot d\bm{S}\,,
\end{equation}
where $d\bm{S}$ is the surface area element. Though physically impractical, the infinite line of mass current along the $z$ axis is employed in this example as a simple and convenient idealization. 

The magnetic fields in these cases can be described as poloidal and toroidal, respectively,  that are naturally interlinked.  To compute the corresponding integral in Eq.~\eqref{S2}, we
 define $x = r \sin\theta \cos\phi$, $y =  r \sin\theta \sin\phi$ and $z = r \cos\theta$, where $(r, \theta, \phi)$ are spherical polar coordinates. The integrand in  Eq.~\eqref{S2} can be expressed as
\begin{equation}\label{S8}
\bar{\bm{A}}_g  \cdot \hat{\bm{B}}_g + \hat{\bm{A}}_g  \cdot \bar{\bm{B}}_g =
\frac{4G^2I_m}{c^2 \sin\theta (r+\zeta)^{n+1}}(W_xJ_x+ W_yJ_y+ W_zJ_z)\,,
\end{equation}
where
\begin{eqnarray}
W_x&=& - \,[ r n\ln(r \sin \theta)\,\sin^2\theta+r+\zeta]\cos\theta \cos\phi\,, \nonumber\\
W_y&=& - \,[ r n \ln(r\sin \theta)\,\sin^2\theta+r+\zeta]\cos\theta \sin\phi\,, \nonumber\\
W_z&=&  \{[ r n\,\sin^2\theta-2(r+\zeta)]\ln(r \sin \theta)+r+\zeta\}\sin\theta\,.\label{S9}
\end{eqnarray}

It is possible to show that $\mathbb{H}_g$, 
\begin{equation}\label{S10}
\mathbb{H}_g = \int (\bar{\bm{A}}_g  \cdot \hat{\bm{B}}_g
+ \hat{\bm{A}}_g  \cdot \bar{\bm{B}}_g)\,r^2 \sin\theta dr d\theta d\phi\,,
\end{equation}
 has finite values starting from $n=4$; indeed, 
\begin{equation}\label{S11}
{\mathbb H}_g = \frac{8\pi G^2}{c^2}C_n\frac{I_mJ_z}{\zeta^{n-3}}\,, 
\end{equation}
where
\begin{equation}\label{S12}
C_n=\frac{8}{(n-1) (n-2) (n-3)}\,.
\end{equation}
The values of $n = 1, 2, 3$ are not allowed because $C_n$ diverges.
To see how this comes about, we first integrate over the azimuthal angle $\phi$ and find
\begin{equation}\label{S13}
{\mathbb H}_g = \frac{8\pi G^2 I_m J_z}{c^2}
\int \frac{W_z}{\sin\theta (r+\zeta)^{n+1}}r^2 \sin\theta dr d\theta\,,
\end{equation}
where $W_z/\sin\theta$ is given by Eq.~\eqref{S9}. Using the integrals given in Appendix~\ref{appint}, Eq.~\eqref{S13} can be evaluated and we obtain Eq.~\eqref{S11} for $\mathbb{H}_g$.

\section{Spacetime Curvature Approach to Gravitomagnetic Helicity}\label{Curvature}

Imagine an arbitrary observer in a gravitational field with spacetime metric $ds^2=g_{\mu \nu} dx^\mu dx^\nu$. To simplify matters, in this section we use units such that $G = c = 1$ throughout. The reference observer follows a future-directed timelike world line $\bar{x}^\mu(\tau)$, where $\tau$ is the observer's proper time. Moreover, the observer carries an orthonormal tetrad frame 
$e^{\mu}{}_{\hat \alpha}$. Here,  $e^{\mu}{}_{\hat 0} = d\bar{x}^\mu/d\tau$ is the observer's 4-velocity vector and 
\begin{equation}\label{F1}
\frac{D e^{\mu}{}_{\hat \alpha}}{d\tau} = \mathbb{F}_{\hat \alpha}{}^{\hat \beta} \,e^{\mu}{}_{\hat \beta}\,,
\end{equation}
where $\mathbb{F}_{\hat \alpha \hat \beta}$ is the observer's acceleration tensor that is antisymmetric due to the tetrad orthonormality condition
\begin{equation}\label{F2}
 g_{\mu \nu} \,e^\mu{}_{\hat \alpha}\,e^\nu{}_{\hat \beta}= \eta_{\hat \alpha \hat \beta}\,. 
\end{equation}
The acceleration tensor can be naturally decomposed into its ``electric" and ``magnetic" parts, namely, $\mathbb{F}_{\hat \alpha \hat \beta} \mapsto (-\bm{g}, \bm{\Omega})$, where $\bm{g}(\tau)$ and $\bm{\Omega}(\tau)$ are spacetime scalars  that represent the translational and rotational accelerations  of the fiducial observer, respectively. Indeed,  $\bm{g}$ has to do with the deviation of reference observer's path $\bar{x}^\mu(\tau)$ from a geodesic and $\bm{\Omega}$ has to do with angular velocity of the rotation of the reference observer's spatial frame with respect to a locally nonrotating (i.e. Fermi-Walker transported) frame. 
 
 For measurement purposes, it proves interesting to set up a quasi-inertial Fermi normal coordinate system in the neighborhood of the fiducial observer.  Imagine all spacelike geodesics that originate from an  event at proper time $\tau$ along the reference world line and are orthogonal to $e^{\mu}{}_{\hat 0}(\tau)$. These geodesics generate a local spacelike hypersurface.   Let $x^\mu$ be an event on this hypersurface such that a \emph{unique} spacelike geodesic connects $x^\mu$ to $\bar{x}^\mu(\tau)$. The Fermi coordinates of $x^\mu$  are defined to be
\begin{equation}\label{F3}
X^{\hat 0} := \tau\,, \qquad X^{\hat i} := \sigma\, \xi^\mu(\tau)\, e_{\mu}{}^{\hat i}(\tau)\,,
\end{equation} 
where $\sigma$ is the proper length of the spacelike geodesic segment from  $x^\mu$ to $\bar{x}^\mu(\tau)$ and $\xi^\mu$, $\xi_\mu(\tau)\,e^{\mu}{}_{\hat 0}(\tau) = 0$, is the unit spacelike vector at $\bar{x}^\mu(\tau)$ that is tangent to the unique  geodesic segment. The reference observer with $X^{\hat \mu} = (\tau, 0, 0, 0)$ is thus fixed at the spatial origin of the Fermi coordinate system. Henceforward, we write Fermi coordinates as $X^{\hat \mu} = (T, \bm{X}) = (T, X, Y, Z)$. The Fermi normal coordinate system is admissible in a cylindrical spacetime region along $\bar{x}^\mu$ with an extent that is characteristic of  the radius of curvature of spacetime along the reference world line. 

The spacetime metric in the Fermi frame is given by
\begin{equation}\label{W1}
ds^2 = g_{\hat \mu \hat \nu}(T, \bm{X})\,dX^{\hat \mu} dX^{\hat \nu}\,,
\end{equation}
where
\begin{equation}\label{W2}
g_{\hat 0 \hat 0} = -\,P^2 + Q^2  - R_{\hat 0 \hat i \hat 0 \hat j}\,X^{\hat i}\,X^{\hat j} + O(|\bm{X}|^3)\,,
\end{equation}
\begin{equation}\label{W3}
g_{\hat 0 \hat i} = Q_{\hat i} -\frac{2}{3} \,R_{\hat 0 \hat j \hat i \hat k}\,X^{\hat j}\,X^{\hat k} + O(|\bm{X}|^3)\,
\end{equation}
and
\begin{equation}\label{W4}
g_{\hat i \hat j} = \delta_{\hat i \hat j} -\frac{1}{3} \,R_{\hat i \hat k \hat j \hat l}\,X^{\hat k}\,X^{\hat l} + O(|\bm{X}|^3)\,,
\end{equation}
where we have neglected third and higher-order terms. Here,  $P$ and $\bm{Q}$,
\begin{equation}\label{W5}
P := 1 + \bm{g}(T) \cdot \bm{X}\,, \qquad \bm{Q} := \bm{\Omega}(T) \times \bm{X}\,,
\end{equation}
are related to the local translational and rotational accelerations of the reference observer, respectively, and
\begin{equation}\label{W6}
R_{\hat \alpha \hat \beta \hat \gamma \hat \delta}(T) = R_{\mu \nu \rho \sigma}\,e^{\mu}{}_{\hat \alpha}\,e^{\nu}{}_{\hat \beta}\,e^{\rho}{}_{\hat \gamma}\,e^{\sigma}{}_{\hat \delta}\,
\end{equation}
is the projection of the Riemann tensor on the tetrad frame of the observer. Taking the symmetries of the Riemann tensor in an arbitrary gravitational field into account, one can express Eq.~\eqref{W6} in the standard manner as a $6\times 6$ matrix with indices that range over the set $\{01,02,03,23,31,12\}$. The general form of this matrix is 
\begin{equation}\label{W7}
\left[\begin{array}{cc}
\mathcal {E} & \mathcal{B}\cr
\mathcal{B}^{\rm T} & \mathcal{S}\cr 
\end{array}\right]\,,
\end{equation}
where $\mathcal{E}$ and $\mathcal{S}$ are symmetric $3\times 3$ matrices and $\mathcal{B}$ is traceless. Here, $\mathcal{E}$, $\mathcal{B}$ and $\mathcal{S}$ represent the measured gravitoelectric,  gravitomagnetic and spatial components of the Riemann curvature tensor, respectively. In close analogy with the treatment in Section \ref{GEM}, we can define the gravitoelectric  potential $\hat{\Phi}$ and gravitomagnetic vector potential $\hat{\bm{A}}$ via $g_{\hat 0 \hat 0} = -\,1 - 2\hat{\Phi}$ and $g_{\hat 0 \hat i} = -\,2\hat{A}_i$. We are particularly interested in the gravitomagnetic aspect; hence,
\begin{equation}\label{W8}
\hat{A}_{\hat i} =  -\,{\frac{1}{2}}Q_{\hat i} + \frac{1}{3}\,R_{\hat 0 \hat j \hat i \hat k}\,X^{\hat j} X^{\hat k}\,.
\end{equation}
Similarly, the corresponding fields can be defined as in Eq.~\eqref{M16}; in fact, to lowest order we find
\begin{equation}\label{W9}
 \hat{B}_{\hat i} = -\,\Omega_{\hat i} -\frac{1}{2}\,\epsilon_{\hat i \hat j \hat k}\,R^{\hat j \hat k}{}_{\hat 0 \hat l}\,X^{\hat l}\,.
\end{equation}
Thus the translational acceleration $\bm{g}$ remains part of the gravitoelectric part and does not influence the gravitomagnetic part. Moreover, we note that $-\,\bm{\Omega}$ is the angular velocity of the   rotation of local ideal gyro axes with respect to the spatial frame of the fiducial observer. The gravitomagnetic terms in Eqs.~\eqref{W8} and~\eqref{W9} depend on the gravitomagnetic components of the curvature tensor; that is, 
\begin{equation}\label{W10}
R_{\hat 0 \hat i}{}^{\hat k \hat l} =  \mathcal{B}_{\hat i \hat j}\,\epsilon^{\hat j \hat k \hat l}\,, \qquad \mathcal{B}_{\hat i \hat j} = \frac{1}{2} \epsilon^{\hat k \hat l}{}_{\hat j}\,R_{\hat 0 \hat i \hat k \hat l}\,.
\end{equation}
We can therefore write Eqs.~\eqref{W8}  and~\eqref{W9} as
\begin{equation}\label{W11}
\hat{A}_{\hat i} =  -\,{\frac{1}{2}}Q_{\hat i} + \frac{1}{3}\,\epsilon_{\hat i \hat k}{}^{\hat l}\,\mathcal{B}_{\hat j \hat l}X^{\hat j} X^{\hat k}\,, \qquad \hat{B}_{\hat i} = -\,\Omega_{\hat i} - \mathcal{B}_{\hat j \hat i}\,X^{\hat j}\,.
\end{equation}

We define the gravitomagnetic helicity in this case as
\begin{equation}\label{W12}
\mathfrak{H} =  \hat{A}_{\hat i} \hat{B}^{\hat i}\,
\end{equation}
and note that Eq.~\eqref{W11} represents the superposition of two fields such that, as we show below, the gravitomagnetic helicity vanishes separately for each field. That is, from Eq.~\eqref{W5}, $\bm{Q}\cdot \bm{\Omega} = 0$; moreover, 
\begin{equation}\label{W13}
\left(\frac{1}{3}\,\epsilon_{\hat i \hat k}{}^{\hat l}\,\mathcal{B}_{\hat s \hat l}X^{\hat s} X^{\hat k}\right)\left(-\mathcal{B}_{\hat j}{}^ {\hat i}X^{\hat j}\right) = -\frac{1}{3}\,\epsilon_{\hat k}{}^{\hat l \hat i}(\mathcal{B}_{\hat s \hat l}X^{\hat s})(\mathcal{B}_{\hat j \hat i}X^{\hat j})\, X^{\hat k} = 0\,.
\end{equation}
Therefore, $\mathfrak{H}$ is given by the cross terms
\begin{equation}\label{W14}
\mathfrak{H} =  \frac{1}{2}Q^{\hat i} \,\mathcal{B}_{\hat j \hat i}X^{\hat j} -\frac{1}{3}\Omega^{\hat i} \,\epsilon_{\hat i \hat k}{}^{\hat l}\,\mathcal{B}_{\hat j \hat l}X^{\hat j} X^{\hat k}\,.
\end{equation}
Using
\begin{equation}\label{W15}
Q_{\hat i}=\epsilon_{\hat i\hat j \hat k}\,\Omega^{\hat j}X^{\hat k}\,,
\end{equation}
we find
\begin{equation}\label{W16}
\mathfrak{H} = \frac{1}{6}\mathcal{B}_{\hat l \hat n}X^{\hat l}Q^{\hat n} =  \frac{1}{6}R_{\hat 0 \hat j \hat i \hat k}\, X^{\hat j}\,\Omega^{\hat i} X^{\hat k}\,.
\end{equation}

The gravitomagnetic helicity is linear in the components of the  $\mathcal{B}$ matrix and quadratic in the spatial Fermi coordinates. Thus, $\mathfrak{H}$ vanishes at the location of the fiducial observer. The coupling  of the rotation of the fiducial observer's frame relative to ideal gyro directions with the measured gravitomagnetic components of the curvature tensor can lead to gravitomagnetic helicity in the Fermi normal coordinate system established around the world line of the reference observer. If the reference observer chooses a Fermi-Walker transported frame, then $\mathfrak{H}$ vanishes. 

It is interesting to evaluate $\mathfrak{H}$ for observers that are spatially at rest in the exterior Kerr spacetime and employ a natural spatial frame with axes that are primarily along the spatial coordinate directions.   

We therefore consider the stationary exterior Kerr spacetime with the metric~\cite{Chandra}
\begin{equation}\label{K1}
ds^2 = -\,dt^2+\frac{\Sigma}{\Delta}dr^2+\Sigma\, d\theta^2 +(r^2+a^2)\sin^2\theta\, d\phi^2
+\frac{2Mr}{\Sigma}(dt-a\sin^2\theta\, d\phi)^2\,,
\end{equation}
where $M$ is the mass of the gravitational source,  $a=J/M$ is the specific angular momentum of the source, $(t,r,\theta,\phi)$ are the standard Boyer-Lindquist coordinates and
\begin{equation}\label{K2}
\Sigma=r^2+a^2\cos^2\theta\,,\qquad \Delta=r^2-2Mr+a^2\,.
\end{equation}

Let us evaluate $\mathfrak{H}$ in Eq.~\eqref{W16} in the case of a Kerr spacetime for the family of static observers that exist in the exterior of the stationary limit surface given by $\Sigma - 2 Mr = \Delta -a^2 \sin^2\theta = 0$. Using the results of Appendix~\ref{appkerr}, we find that $\mathfrak{H}$ can be expressed as
\begin{equation}\label{K3}
\mathfrak{H} = -\,{\frac{1}{2}} M^2a^2\,\frac{\Delta\,X\,\cos\theta
+ r\,\Delta^{1/2}\,Y\, \sin \theta}{\Sigma^{5/2} (\Sigma-2Mr)^2}\,Z\,\sin \theta\,,
\end{equation}
where the spherical polar coordinates $(r, \theta, \phi)$ characterize the fixed spatial position of the observer. The exterior Kerr spacetime is stationary; therefore, gravitomagnetic helicity is time independent.    

It is interesting to note that gravitomagnetic helicity in Kerr spacetime is proportional to $J^2$, namely, the square of the angular momentum of the source, and vanishes along the Kerr rotation axis as well as  for $Z = 0$.

\section{Discussion}\label{Discussion}

We have studied gravitomagnetic helicity within both approaches to GEM.  While the linear perturbation approach to GEM emphasizes the analogy with magnetic helicity, the spacetime curvature approach leads to a result that is linearly proportional to the gravitomagnetic components of the curvature tensor. We have used the latter approach to calculate the gravitomagnetic helicity $\mathfrak{H}$ in the Kerr field within a Fermi normal coordinate system established along the world line of an arbitrary observer that is spatially at rest and uses a natural orthonormal tetrad frame. The Kerr gravitomagnetic helicity crucially depends on the coupling of the rotation of the observer's frame with the gravitomagnetic components of the Riemann curvature tensor. To illustrate this point, we examine the two components of this coupling in turn.

 A nonzero $\mathfrak{H}$ depends on the rotation of the fiducial observer's local reference frame relative to ideal nonrotating (i.e. Fermi-Walker transported) axes. Therefore, consider the congruence of reference observers at rest in any spacetime  that is conformally flat, i.e. $g_{\mu \nu}(x^\alpha)  = \mathbb{C}^2(x^\alpha)\,\eta_{\mu \nu}$. The axes of the natural tetrad frame of these observers point along the Cartesian coordinate axes $x^\alpha = (t, x, y, z)$. It is intuitively clear that conformal scaling does not induce a rotation in this case; that is,  $\bm{\Omega} = 0$. Indeed, one can show that in this simple spacetime geometry  $\mathfrak{H} = 0$. A similar result is obtained if the gravitomagnetic components of the Riemann curvature tensor vanish, which is the case for the G\"odel spacetime. That is, one can show that for observers that are spatially at rest in this rotating universe, the gravitomagnetic components of the spacetime curvature tensor vanish~\cite{ChMa}; therefore, the corresponding gravitomagnetic helicity $\mathfrak{H}$ vanishes for the G\"odel universe as well. 
 
In connection with the possibility of measurement of gravitomagnetic helicity $\mathfrak{H}$, it is interesting to determine this quantity for observers that are spatially at rest in the recently constructed linearized gravitational radiation field that carries orbital angular momentum~\cite{Iwo}. This is, however,  beyond the scope of the present investigation.

\section*{Acknowledgments}

DB is grateful to R. T. Jantzen and O. Semer\'ak for helpful comments. The work of YNO was supported in part by the Russian Foundation for Basic Research (Grant No. 18-02-40056-mega).

\appendix

\section{Fermion in gravitomagnetic field}\label{appdirac}

The Dirac matrices $\beta, \bm{\alpha}$ and the spin matrix $\bm{\Sigma}$ read explicitly
\begin{equation}\label{albe}
\beta = \left(\begin{array}{cc}\mathbb{I} & 0\\ 0 & -\mathbb{I}\end{array}\right)\,,\quad
\bm{\alpha} = \left(\begin{array}{cc} 0 & \bm{\sigma}\\ \bm{\sigma}&0\end{array}\right)\,,
\quad \bm{\Sigma} = \left(\begin{array}{cc}\bm{\sigma} & 0\\ 0 & \bm{\sigma}\end{array}\right)\,,
\end{equation}
where $\mathbb{I}$ is the $2\times 2$ unit matrix and $\bm{\sigma} = (\sigma_x, \sigma_y, \sigma_z)$ are the standard Pauli matrices. In cylindrical coordinates (\ref{L4}), we find
\begin{equation}\label{sigp}
\sigma_x\widehat{p}_x + \sigma_y\widehat{p}_y = - \left(\begin{array}{cc} 0 & e^{-i\varphi}\\ e^{i\varphi} & 0
\end{array}\right) i\hbar\,\partial_\varrho + \left(\begin{array}{cc} 0 & - e^{-i\varphi}\\ e^{i\varphi} & 0
\end{array}\right) {\frac \hbar \varrho}\,\partial_\varphi\,. 
\end{equation}

Exact solutions for a fermion in the gravitomagnetic field can be found by plugging ansatz (\ref{psi}) into the Schr\"odinger equation (\ref{Sch}). As usual, the Dirac theory admits solutions with the energy of both signs. A direct computation yields, respectively:
\begin{align}
{\mathit \Psi}{}_{+}(t,\varrho,\varphi,z) &= e^{-i\left[|\omega_0| - (n + {\frac s2})B_g/c\right]t + i(k_zz + n\varphi)}
\,{\mathcal U}^{(s)}_{n,k_z,\eta}\,,\label{sP}\\
{\mathit \Psi}{}_{-}(t,\varrho,\varphi,z) &= e^{i\left[|\omega_0| + (n + {\frac s2})B_g/c\right]t + i(k_zz - n\varphi)}
\,{\mathcal V}^{(s)}_{n,k_z,\eta}\,,\label{sN}
\end{align}
where $\omega_0$ is given by (\ref{om0}), $n = 0, \pm 1, \pm 2, \dots$, the spin variable $s = \pm 1$, and the 4-spinors read (denoting $\omega'_0 = |\omega_0| + mc^2/\hbar$)
\begin{align}
{\mathcal U}^{(+1)}_{n,k_z,\eta} &= N_0
\left(\begin{array}{c} J_n(\eta\varrho) \\ 0 \\ {\frac {ck_z}{\omega'_0}}\,J_n(\eta\varrho) \\
{\frac {ic\eta}{\omega'_0}}\,e^{i\varphi}J_{n+1}(\eta\varrho) \end{array}\right)\,,\quad
{\mathcal U}^{(-1)}_{n,k_z,\eta} = N_0
\left(\begin{array}{c} 0 \\ \,J_n(\eta\varrho) \\ {\frac {-ic\eta}{\omega'_0}}\,e^{-i\varphi}J_{n-1}(\eta\varrho)\\
{\frac {-\,ck_z}{\omega'_0}}\,J_n(\eta\varrho) \end{array}\right)\,,\label{U}\\
{\mathcal V}^{(+1)}_{n,k_z,\eta} &= N_0
\left(\begin{array}{c} {\frac {ck_z}{\omega'_0}}\,J_n(\eta\varrho) \\ {\frac {-\,ic\eta}
{\omega'_0}}\,e^{i\varphi}J_{n+1}(\eta\varrho) \\ J_n(\eta\varrho) \\ 0 \end{array}\right)\,,\quad 
{\mathcal V}^{(-1)}_{n,k_z,\eta} = N_0
\left(\begin{array}{c} {\frac {ic\eta}{\omega'_0}}\,e^{-i\varphi}J_{n-1}(\eta\varrho) \\
{\frac {-\,ck_z}{\omega'_0}}\,J_n(\eta\varrho)\\ 0 \\ J_n(\eta\varrho) \end{array}\right)\,.\label{V}
\end{align}
The normalization constant can be chosen as $N_0^2 = {\frac {\omega'_0}{2|\omega_0|}}$.

Recalling the definition of the charge conjugation by ${\mathit \Psi}{}^c := C\overline{{\mathit \Psi}{}}{\,}^{\rm T}$, where $C = -\,i\alpha^{\hat{2}}$, one can check that the spinors above are charge-conjugated as follows:
\begin{equation}\label{ChC}
({\mathcal U}^{(s)}_{n,k_z,\eta}){}^c = s\,{\mathcal V}^{(-s)}_{-n,k_z,\eta}\,.
\end{equation}

\section{Useful Integrals}\label{appint}

In Section \ref{Helicity}, the integrals that we need to evaluate gravitomagnetic helicity~\eqref{S13} are given below.

We start with
\begin{equation}\label{A1}
 \int _{0}^{\pi} \ln( \sin \theta) \sin \theta d\theta = -\,2 + 2\ln 2\,, 
\end{equation}
\begin{equation}\label{A2}
\int _{0}^{\pi} \ln( \sin \theta) \sin^3 \theta d\theta = -\,{\frac{10}{9}} + \frac{4}{3}\ln 2\,.
\end{equation}
Moreover, it is useful to introduce $N$, 
\begin{equation}\label{A3}
 N := (n-1) (n-2) (n-3)\,, 
\end{equation}
for the sake of simplicity; then, 
\begin{equation}\label{A4}
\int_{0}^{\infty} \frac{r^2}{(r+\zeta)^n} \,dr = \frac{2}{\zeta^{n-3} N}\,,
\end{equation}
\begin{equation}\label{A5}
\int_{0}^{\infty} \ln r \,\frac{r^2}{(r+\zeta)^n} \,dr= - \,2\frac{\psi_0(n) +\gamma_0
-\ln\zeta}{\zeta^{n-3} N}+\frac{4+9 n-12 n^2+3 n^3}{\zeta^{n-3} N^2}\,, 
\end{equation}
\begin{equation}\label{A6}
\int _{0}^{\infty} \ln r \,\frac{r^3}{(r+\zeta)^{n+1}}\, dr = -\,6\,\frac{\psi_0(n)
+ \gamma_0-\ln\zeta}{\zeta^{n-3}\,n\,N}+\frac{49 -48 n+11 n^2}{\zeta^{n-3}\,N^2}\,.
\end{equation}
Here, $\psi_0(x)$ denotes the digamma function
\begin{equation}\label{A6a}
\psi_0(x) := \frac{d}{dx}\ln \Gamma(x) = \frac{1}{\Gamma(x)}\frac{d}{dx} \Gamma(x)\,.  
\end{equation}
For positive integer values of $x$, $\psi_0(x)$ is given by
\begin{equation}\label{A7}
\psi_0(n) =  -\,\gamma_0 + \sum_{k = 1}^{n-1} \frac{1}{k}\,, 
\end{equation}
where $\psi_0(1) = -\,\gamma_0$ and  $\gamma_0 :=0.5772156649...$ is the Euler-Mascheroni constant.

\section{Curvature of the Kerr Field}\label{appkerr}

We are interested in the curvature of the Kerr field as measured by the \emph{static}  family of accelerated observers  with the adapted orthonormal frame
\begin{align}\label{B1}
\nonumber e_{\hat 0}={}& (-g_{tt})^{-1/2}\,\partial_t\,,  \qquad e_{\hat 1}=(g_{rr})^{-1/2}\, \partial_r\,, \qquad  e_{\hat 2}= (g_{\theta\theta})^{-1/2}\, \partial_\theta\,,\\
 e_{\hat 3}={}& \left(g_{\phi\phi}-\frac{g_{t\phi}^2}{g_{tt}}\right)^{-1/2}\,\left(-\frac{g_{t\phi}}{g_{tt}} \,\partial_t+\partial_\phi\right)\,,
\end{align}
where the tetrad axes are primarily along the  Boyer-Lindquist coordinate directions. The explicit expressions for the Kerr potentials are given in Eq.~\eqref{K1}; therefore,  
\begin{align}\label{B2}
\nonumber e_{\hat 0}={}&\left(\frac{\Sigma}{\Sigma - 2 Mr}\right)^{1/2} \partial_t\,,  \qquad e_{\hat 1}= \left(\frac{\Delta}{\Sigma}\right)^{1/2}\, \partial_r\,, \qquad  e_{\hat 2}=\left(\frac{1}{\Sigma}\right)^{1/2}\,\partial_\theta\,,\\
e_{\hat 3} ={}&  -2Ma\,\frac{r \sin \theta}{[\Delta \Sigma (\Sigma - 2 Mr)]^{1/2}}\, \partial_t +\left(\frac{\Sigma -2Mr}{\Delta\,\Sigma}\right)^{1/2}\,\frac{1}{\sin \theta}\, \partial_\phi\,.
\end{align}
We can now use Eq.~\eqref{F1} to evaluate $\bm{g}$ and $\bm{\Omega}$. The results are
\begin{eqnarray}\label{B3}
g^{\hat i}\,e_{\hat i} &=& \frac{M}{\Sigma^{3/2}(\Sigma-2Mr)}\left[\Delta^{1/2}\,(r^2 -a^2 \cos^2\theta)
\, e_{\hat 1} - 2  r a^2\sin \theta \cos \theta\,  e_{\hat 2}\right]\,,\\
\Omega^{\hat i}\,e_{\hat i} &=& -\,{\frac{Ma}{\Sigma^{3/2}(\Sigma-2Mr)}}\left[2\Delta^{1/2}\,
r \cos\theta \,e_{\hat 1} + (r^2-a^2\cos^2\theta)\,\sin \theta\,e_{\hat 2}\right]\,.\label{B4}
\end{eqnarray}

Kerr spacetime is Ricci flat, therefore,  Eq.~\eqref{W7} for static observers takes the form
\begin{equation}\label{B5}
\left[\begin{array}{cc}
\mathcal{E} & \mathcal{B}\cr
\mathcal{B} & -\mathcal{E}\cr 
\end{array}\right]\,,
\end{equation}
where  $\mathcal{E}$ and $\mathcal{B}$ are now symmetric and traceless. The Riemann curvature tensor for $R_{\mu \nu } = 0$ degenerates  into the Weyl conformal curvature tensor $C_{\mu\nu\rho\sigma}$ whose gravitoelectric and gravitomagnetic components are then 
\begin{equation}\label{B6} 
\mathcal{E}_{\hat a \hat b}= C_{\alpha\beta\gamma\delta}\,e^\alpha{}_{\hat 0}\, e^\beta{}_{\hat a}
\, e^\gamma{}_{\hat 0}\, e^\delta{}_{\hat b}\,,\qquad \mathcal{B}_{\hat a \hat b} =
C^*_{\alpha\beta\gamma\delta}\,e^\alpha{}_{\hat 0}\,e^\beta{}_{\hat a}\, e^\gamma{}_{\hat 0}\, e^\delta{}_{\hat b}\,,
\end{equation}
where  $C^*_{\alpha\beta\gamma\delta}$ is the unique dual of the Weyl tensor given by
\begin{equation}\label{B7} 
C^*_{\alpha\beta\gamma\delta}=\frac12\, \epsilon_{\alpha \beta}{}^{\mu\nu}\,C_{\mu\nu\gamma\delta}\,,
\end{equation}
since the right and left duals of the Weyl  tensor coincide. Here, $\epsilon_{\alpha \beta \gamma \delta}$ is the Levi-Civita tensor; in our convention, $\epsilon_{\hat 0 \hat 1 \hat 2 \hat 3} = 1$ and $\epsilon_{\hat 0 \hat i \hat j \hat k} = \epsilon_{\hat i \hat j \hat k}$.

With respect to the static observers, the nonvanishing components of the tidal matrix are given by~\cite{Bini:2016xqg} 
\begin{eqnarray}\label{B8}
\mathcal {E}_{\hat 1 \hat 1}&=& -\,2\mathbb{E}\,\frac{\Delta +\frac{1}{2}\,a^2 \sin^2 \theta}{\Delta -a^2\sin^2 \theta}\,,\nonumber\\
\mathcal {E}_{\hat 1 \hat 2}&=& -\,3a\,\sin \theta\, \mathbb{B} \,\frac{\Delta^{1/2}}{\Delta-a^2\,\sin^2 \theta}\,,\nonumber\\
\mathcal {E}_{\hat 2 \hat 2}&=&  \mathbb{E} \,\frac{\Delta + 2\,a^2 \sin^2 \theta}{\Delta -a^2\sin^2 \theta}\,,\nonumber\\
{\mathcal E}_{\hat 3 \hat 3}&=& \mathbb{E}\,,
\end{eqnarray}
where
\begin{equation}\label{B9}
 \mathbb{E} = \frac{M r (r^2-3 a^2\cos^2 \theta)}{\Sigma^3}\,, \qquad 
 \mathbb{B}= -\,\frac{M a (3r^2- a^2\cos^2 \theta)\,\cos \theta }{\Sigma^3}\,.
\end{equation}
Moreover, the nonzero elements of the gravitomagnetic part of the Weyl curvature are given by
\begin{eqnarray}\label{B10}
\mathcal {B}_{\hat 1 \hat 1}&=& -\,2 \mathbb{B} \,\frac{\Delta +\frac{1}{2}\,a^2 \sin^2 \theta}{\Delta -a^2\sin^2 \theta}\,,\nonumber\\
\mathcal {B}_{\hat 1 \hat 2}&=& 3 a\,\sin \theta \,\mathbb{E} \,\frac{\Delta^{1/2}}{\Delta-a^2\,\sin^2 \theta}\,,\nonumber\\
\mathcal {B}_{\hat 2 \hat 2}&=&  \mathbb{B} \,\frac{\Delta + 2\,a^2 \sin^2 \theta}{\Delta -a^2\sin^2 \theta}\,,\nonumber\\
{\mathcal B}_{\hat 3 \hat 3}&=& \mathbb{B}\,.
\end{eqnarray}

The Kerr field is of type D in the Petrov classification and this circumstance accounts for a certain ``parallelism" that is evident between the gravitoelectric and gravitomagnetic components of its curvature.

\end{document}